\advance\hoffset -0.5cm
\advance\voffset 0cm
\documentstyle[12pt]{article} 

\newlength{\height}
\divide \height by 3

\begin{document} 
\newcommand{\be}{\begin{equation}}
\newcommand{\ee}{\end{equation}}
\newcommand{\bea}{\begin{eqnarray}}
\newcommand{\eea}{\end{eqnarray}}
\newcommand{\req}[1]{(\ref{#1})}
\def\spur#1{\mathord{\not\mathrel{#1}}}
\baselineskip=\height 
\begin{titlepage}
\begin{center}
\makebox[\textwidth][r]{SNUTP 92-92}
\vskip 0.35in
{{\Large \bf $\theta$ EFFECTS IN CHERN-SIMONS ${\rm QED}_{2+1}$\\
WITH A FOUR-FERMI INTERACTION}}
\end{center}
\begin{center}
  \par \vskip .1in \noindent
Junegone Chay$^1$
Deog Ki Hong$^2$,
Taejin Lee$^3$, S.~H.~Park$^4$
\par \vskip .1in \noindent
\end{center}

\begin{center}
$^1$Department of Physics, Korea University \\
Seoul 136-701, Korea\\

\par \vskip .1in \noindent

$^2$Department of Physics, Pusan National University\\
Pusan 609-735, Korea\\

\par \vskip .1in \noindent

$^3$Department of Physics, Kangwon National University\\
Chunchon 200-701, Korea\\

\par \vskip .1in \noindent

$^4$Center for Theoretical Physics, Department of Physics\\
Seoul National University, Seoul, Korea
\par \vskip .1in \noindent
\end{center}

\begin{abstract}
We investigate the effects of the Chern-Simons coupling on the high
energy behavior in the $(2+1)$-dimensional Chern-Simons QED with a
four-Fermi interaction. Using the $1/N$ expansion we discuss the
Chern-Simons effects on the critical four-Fermi coupling at $O(1/N)$
and the $\beta$ function around it. High-energy behavior of Green's
functions is also discussed. By explicit calculation, we find that the
radiative correction to the Chern-Simons coupling vanishes at $O(1/N)$
in the broken phase of the dynamical parity symmetry. We argue that no
radiative corrections to the Chern-Simons term arise at higher orders
in the $1/N$ expansion.
\end{abstract}
\end{titlepage}
Various phenomena peculiar in (2+1) dimensions happen when the
Chern-Simons (CS) term \cite{schon,hagen} is present. Combined with
the ordinary Maxwell term, the CS term generates a gauge invariant
mass for gauge fields. There has been a great interest in the
so-called CS theory where the kinetic action for gauge fields is
characterized only by the CS term \cite{hagen}. In this case the gauge
sector of the theory is renormalizable (though not super-renormalizable)
in (2+1) dimensions, which makes it field theoretically interesting.
While the CS term has been known to affect the long-distance behavior
of the theory, our motivation in this paper is to study the CS effects
on the short-distance behavior.

In this paper we investigate the effects of the CS term on the high
energy behavior of the CS ${\rm QED}_{2+1}$ with a four-Fermi interaction.
It has been shown \cite{rosen} that a class of theories with a four-Fermi
interaction is renormalizable in the framework of the $1/N$ expansion
in spite of its non-renormalizability in the weak coupling expansion.
The $1/N$ technique also shows various nonperturbative phenomena among
which is the interesting renormalization flow for the four-Fermi
coupling. At leading order in $1/N$, the four-Fermi
coupling has a nontrivial ultraviolet fixed point which survives
beyond leading order in $1/N$.

We find the dependence of the $\beta$ function on $\theta$ (the
coupling of the CS term) for the four-Fermi coupling at $O({1/N})$.
The correction to the critical coupling is calculated.
The nature of the ultraviolet fixed point at leading order
strongly depends on $\theta$. It is drastically different from the
model with the usual Maxwell term which has been previously
considered in connection with the dynamical symmetry pattern and
the critical behavior of the theory \cite{semone}.

In this model another interesting issue is the radiative
correction to $\theta$ which has been frequently discussed in the
(2+1)-dimensional gauge theories
\cite{coleman} - \cite{hong}. By explicit
calculation, we show that $\theta$ does not renormalize at $O(1/N)$.
We also prove that there are no radiative corrections to $\theta$ at
all orders in $1/N$, thus extend the non-renormalization theorem by
CoBleman and Hill \cite{coleman} to the $1/N$ expansion.

The (2+1)-dimensional CS QED with the simplest four-Fermi
interaction is given in the Euclidean version by
\be
{\cal L} = i {\overline \psi}_j \spur{D} \psi_j + {g^2 \over 2N}
\left({\overline \psi}_j \psi_j \right)^2 + i \theta \epsilon_{\mu
\nu \rho} A_{\mu} F_{\nu \rho},
\label{lagrangian}
\ee
where $D_\mu = \partial_\mu + ieA_{\mu}/\sqrt{N}$ and $j$ is summed
over from 1 to $N$. The couplings $e$ and $\theta$ are dimensionless
and $\psi_j$ are two-component spinors. The $\gamma$ matrices are
defined as
\be
\gamma_1 = \sigma^3,~ \gamma_2 = \sigma^1,~ \gamma_3= \sigma^2.
\label{gammas}
\ee

Introducing an auxiliary field $\sigma$ to facilitate the $1/N$
expansion \cite{rosen}, we can rewrite Eq.~(\ref{lagrangian}) as
\be
{\cal L} = i {\overline \psi}_j \spur{D} \psi_j
+ i \sigma {\overline \psi}_j
\psi_j + {N \sigma^2 \over 2 g^2} + i \theta
\epsilon_{\mu \nu \rho} A_{\mu} F_{\nu \rho}
\label{sigmalag}
\ee
At leading order in $1/N$, the theory has a two-phase structure
characterized by the order parameter $\left< \sigma \right>$ as in the
case without the gauge field \cite{rosen}. For the weak coupling,
$g^2 < g_c{}^2$, where $g_c{}^{-2} = 2 \int \! {d^3 p \over (2 \pi)^3}
{1 \over p^2}$, the parity is unbroken characterized by $\left< \sigma
\right>= 0$. If the coupling is larger than $g_c^2$, the auxiliary
field $\sigma$ gets a non-zero vacuum expectation value. The parity
symmetry is dynamically broken and the fermion acquires the mass equal
to $\left< \sigma \right> = M$. We consider this broken phase in this
paper.

At leading order in $1/N$, the vacuum satisfies the following relation
\be
{1 \over g^2} - 2 \int_{\Lambda} \! {d^3 p \over (2 \pi)^3}
{1 \over p^2 + M^2} = 0,
\label{criticalg}
\ee
where we introduce the high-momentum cutoff $\Lambda$. We can easily see
from Eq.~(\ref{criticalg}) that the dimensionless charge defined by
$\lambda = 1 / (g^2 \Lambda)$ flows to the finite value
$\lambda_c = 1 / \pi^2$ in the continuum limit $(\Lambda \rightarrow
\infty)$. The $\beta$-function for $\lambda$ in the vicinity of
$\lambda_c$, at leading order in $N$, is $\beta (\lambda) =
- (\lambda - \lambda_c)$ \cite{rosen}.

The Feynman diagrams in the broken phase are depicted in Fig.~1. The
propagator for the fermion $\psi$ is given by
\be
S(p) = \frac{1}{\spur{p} + iM},
\label{fermion}
\ee
the propagator for the auxiliary field $\sigma$ is
\be \displaystyle
D(p^2) = {{-4 \pi \sqrt {p^2}} \over {(p^2 - 4 M^2) \tan^{-1}
{{\sqrt{p^2}} \over {2M}}}}.
\label{auxil}
\ee
The photon propagator is given by
\be
G_{\mu\nu} = {1 \over p^2} \left( \delta_{\mu\nu} - {p_\mu p_\nu \over
p^2} \right) \Pi_1 (p^2) + \epsilon_{\mu \nu \rho} {p_\rho \over p^2}
\Pi_2 (p^2),
\label{gmunu}
\ee
where $\Pi_1$ and $\Pi_2$ are given by
\be
\Pi_1 (p^2) = {\Pi_e \over \Pi_e{}^2/p^2 + (\Pi_o + \theta)^2}
\label{pione}
\ee
\be
\Pi_2 (p^2) = {(\Pi_o + \theta) \over \Pi_e{}^2/p^2 +
(\Pi_o + \theta)^2},
\label{pitwo}
\ee
The resummation technique of the $1/N$ expansion results in
the photon propagator and
$\Pi_e$ [$\Pi_o$] in Eq.~(\ref{pione}) [(\ref{pitwo})] is the
even (odd) part of the vacuum polarization:
\be
\Pi_e (p^2) = {e^2 \over 8 \pi} \left( 2M + {p^2 - 4 M^2 \over
\sqrt {p^2}} {\rm tan}^{-1} {\sqrt {p^2} \over 2M} \right)
\label{pieven}
\ee
and
\be
\Pi_o (p^2) = {Me^2 \over 2 \pi} {1 \over \sqrt {p^2}}
{\rm tan}^{-1} {\sqrt {p^2} \over 2M}.
\label{piodd}
\ee
As $p \rightarrow \infty$, the photon
propagator behaves like
${\kappa \over \kappa^2 + \theta^2} p^{-1}$ (the symmetric part) and
${\theta \over \kappa^2 + \theta^2} p^{-1}$ (the antisymmetric part),
where $\kappa = e^2/16$.

The vertex of $\sigma {\overline \psi}\psi$ is given by
$-i\delta_{ij}/\sqrt{N}$ and the vertex of $A_{\mu}
{\overline \psi}\psi$ is given by $-e\gamma_{\mu}
\delta_{ij}/\sqrt{N}$. Notice that the graphs in Fig.~2 are forbidden
to avoid double counting.

The renormalizability of the theory can be easily proved using a
simple power counting and the Ward identity. We refer to
Ref.~\cite{rosen} for detailed renormalization procedure.
In terms of the renormalized quantities, the Lagrangian density
(See Eq.~(\ref{sigmalag}).) can be rewritten as
\be
{\cal L} = i Z_1 {\overline \psi}_j \spur{D} \psi_j + i Z_2 \sigma
{\overline  \psi}_j \psi_j +
{N Z_3 \over 2 g^2} {Z_2{}^2 \over Z_1{}^2} \sigma^2
+ i Z_4 \theta \epsilon_{\mu \nu \rho} A_{\mu} F_{\nu \rho}
\label{finallag}
\ee
Note that the Maxwell term is not generated as a counter term, which
assures the renormalizability of the theory. We keep
$g^2$ as in Eq.~(\ref{criticalg}), so that we are in the broken phase.
Expanding $\sigma(x) =
(Z_1/Z_2) (M + \sigma'(x)/\sqrt N)$ in powers of $1/N$, we have at
next-to-leading order
\be
{Z_2 \over Z_1} \left<\sigma(x)\right> = M + {1 \over \sqrt N}
\left<\sigma'(x)\right>,
\ee
where
\be
{\sqrt N  \left<\sigma^{\prime}(x)\right> \over D(0)} =
- {M Z_3^{\prime} \over g^2} + {\rm Tadpole}\ {\rm Diagrams}\ {\rm in}\
{\rm Fig.~3(c)},
\label{sigmaprime}
\ee
with $Z_3 = 1 + Z_3^{\prime}/N$.

In order to calculate the renormalization constants $Z_i$, we evaluate
the diagrams in Fig.~3. We have
\be
Z_1 = 1 - {4 \over {3\pi^2N}} \left( 1 + {{4\kappa^2} \over {\kappa^2 +
\theta^2}} \right) \ln ({{\Lambda} \over {\mu}}),
\label{zone}
\ee

\be
Z_2 = 1 + {4 \over {\pi^2N}} \left( 1 + {{12\kappa^2} \over
{\kappa^2 + \theta^2}} \right) \ln ({{\Lambda} \over {\mu}}),
\label{ztwo}
\ee
and
\bea
{Z_3 \over g^2} &=& {1\over \pi^2} \Big[ 1 - {2\over N}
{{\theta^2 - \kappa^2} \over {\kappa^2 + \theta^2}}\Big] \Lambda
\nonumber \\
&& [B[B- {M \over {2 \pi}} \Big[1 + {{16} \over {3\pi^2N}}
\left(1 + {{10\kappa^2} \over {\kappa^2 + \theta^2}} \right)
\ln ({{\Lambda}\over {\mu}})\Big].
\label{zthree}
\eea
Note that there is no dangerous $\Lambda {\rm ln}\Lambda$ dependence
in Eq.~(\ref{zthree}), which preserves the phase structure at leading
order. The $\Lambda^2$ divergence from the tadpole diagram with
internal photon line in Fig.~3(c) only shifts the gap
equation and does not affect the phase structure of the theory.

The extra parameter $\mu$ in the above equations is an unphysical
renormalization point. The $\mu$ dependence in Eq.~(\ref{zthree}) can
be absorbed in $M_{\rm phys}$, then $Z_3$ is rewritten as
\be
{Z_3 \over g^2 \Lambda} = {1 \over {\pi^2}} \left( 1 +
{2 \over N}{{1-x^2} \over {1 + x^2}} \right) - {1 \over 2 \pi}
\left( M_{\rm phys}/\Lambda \right)^A
\label{newzthree}
\ee
where
\be
A = 1 - {16 \over 3 \pi^2}{1 \over N} \left( 1 + {{10} \over {1 + x^2}}
\right) > 0
\label{powera}
\ee
with $x=\theta /\kappa$. After rescaling $A_{\mu} \rightarrow
A_\mu/e$, the theory has one gauge coupling $x$.
The exact $S$-matrix depends on just one scale $M_{\rm phys}$,
so the invariant charge $Z_3 / g^2$ depends on $\Lambda$ and
$M_{\rm phys}$. The leading-order parameter $M$ is no longer equal to
$M_{\rm phys}$, and sholud be regarded as a function of $\mu$.

The $\beta$ function for $\lambda = Z_3/(g^2 \Lambda)$ around
$\lambda_c$ is given by
\be
\beta(\lambda) = - A \left(\lambda - \lambda_c \right),
\label{betaflow}
\ee
where
\be
\lambda_c = {1 \over \pi^2} \left(1 + {2 \over N}
{{1-x^2}\over {1+x^2}}\right).
\label{lambdac}
\ee
The finite ultraviolet fixed point still exists. When $x = 1$, it is
not shifted at $O(1/N)$. It moves downwards (upwards) when $x > (<)
1$. The slope of the $\beta$ function for $\lambda$ becomes $-A$.
In the limit $x \rightarrow \infty$, the gauge degrees of freedom drop
out and the theory has only the four-Fermi interaction. This can be
immediately checked by taking the limit $x \rightarrow \infty$ in
the Eqs.~(\ref{powera}) and (\ref{lambdac}) and comparing them with
the previous results in the four-Fermi interaction model \cite{rosen}.
In this limit, $A$ is always
positive even when $N=1$. Therefore the theory is consistent for any $N$.
As we can see from Fig.~4 in which we plot $x$ versus $N$,
$N$ should be larger than 6 to make the theory consistent for any
value of $x$.

{}From the renormalization constants in Eqs.~(\ref{zone}) and
(\ref{ztwo}) we can easily read
the ultraviolet dimension of the fields $\psi$ and $\sigma$
\be
\left[\psi\right] = 1 + {2\over {3\pi^2N}} \left( 1+ {4\over {1+x^2}}
\right),
\label{psidim}
\ee
\be
\left[\sigma\right] = A.
\ee
Then the high energy behavior of the connected, truncated Green's
function with $n$ external fermion legs and $m$ external $\sigma$ legs
is $\sim E^p$ where $p=3 - n\left[\psi\right] - m\left[\sigma\right]$.
If the Maxwell term exists in the original theory, which we do not
cosider here, the QED sector is finite, thus it does not influence the
high-energy behavior of the theory which is characterized by the
four-Fermion interaction \cite{rosen}.

We now discuss the radiative correction to $\theta$ at $O(1/N)$,
which is given by $\Pi_o(0)=\lim_{p\rightarrow 0}{1\over
6}\epsilon^{\mu\nu\lambda} {\partial\over\partial p^\lambda}\Pi_{\mu\nu}
(p)$. The leading order corrections to the vacuum polarization
$\Pi_{\mu\nu}$ can be calculated from the diagram in
Fig.~2 as in the weak coupling expansion. This is summed to
the photon propagator in Eqs.~(\ref{gmunu})--(\ref{piodd}). {}From
Eq.~(\ref{piodd}) its contribution
to $\Pi_o(0)$ is $- e^2/4\pi$ as is well known.

The corrections at $O(1/N)$ arise from the integrals depicted in the
diagrams of Figs.~5 and 6. After taking the trace of the $\gamma$
matrices in the integrands, we find that there are no contributions to
$\Pi_o(0)$ from the Feynman diagrams of Figs.~5(a).
The diagrams Figs.~5(b) have been already
discussed in the contexts of the Maxwell-Chern-Simons QED \cite{bern}
and the Chern-Simons QED \cite{semtwo,hong} in the weak coupling
expansion scheme. Recalling the
explicit calculations in Refs.~\cite{bern,semtwo}, we can deduce that
they also have null contribution to the coefficient of the CS term. We
may write their contributions to $\Pi_o(0)$ by
\be
\Delta_1\Pi_o(0) = \lim_{p\rightarrow 0}{1\over
6}\epsilon^{\mu\nu\lambda} {\partial\over\partial
p^\lambda}\Delta_1\Pi_{\mu\nu} (p)
\label{deltapio}
\ee
\be
\Delta_1\Pi_{\mu\nu} (p) = \int \! {d^3 q\over (2\pi)^3}
\Gamma_{\mu\nu\rho\sigma}(p,-p, q,-q) G^{\rho\sigma}(q),
\label{deltapi}
\ee
where $\Gamma_{\mu\nu\rho\sigma}$ is the one-loop 4-photon function.

The calculations in Ref.~6 show that the same expression for
$\Delta_1\Pi_o(0)$ exactly vanishes when the photon propagator in the
Maxwell-Chern-Simons QED is given by
\be
G_{\rho\sigma}(q) = G_{\rho\sigma}^e(q)+G_{\rho\sigma}^o(q),
\ee
where
\be
 G_{\rho\sigma}^e(q) = {1 \over q^2 + \theta^2}\left(\delta_{\rho\sigma}
-{q_\rho q_\sigma\over q^2}\right),
\ee
\be
G_{\rho\sigma}^o(q) = {\theta \over
q^2 + \theta^2}\epsilon_{\rho\sigma\lambda}{q^\lambda\over q^2}.
\ee
In fact, the contributions to
$\Delta_1\Pi_o(0)$ with $G_{\rho\sigma}^e(q)$ and $G_{\rho\sigma}^o(q)$
separately vanish even
before the gauge-field loop integration is performed. Since the  induced
photon propagator Eq.~(\ref{gmunu}) which must be employed in
Eqs.~(\ref{deltapio}) and (\ref{deltapi}) has
the same structure, it becomes clear that $\Delta_1\Pi_o(0)$ exactly
vanishes.

The diagrams in Fig.~6 are of three loop order in the weak coupling
expansion, but produce the same $O(1/N)$ corrections as the diagrams in
Fig.~5. We denote the one-loop $m$-photon $n$-$\sigma$
function by $\Gamma_{\mu_1\dots\mu_m (n)}$.
The corrections from the diagrams of Fig.~6(a) and those of
Fig.~6(b) involve the one-loop 2-photon, 1-$\sigma$ field
functions $\Gamma_{\mu\nu (1)}$ which is given by
\bea
\Gamma_{\mu\nu (1)}(p,-q,q-p) &=& - {e^2 \over \sqrt N} \int \!
{d^3 k\over (2\pi)^3}{\rm tr} \Bigl(\gamma_{\mu} S(k-p)
\gamma_{\nu} S(k+q-p) \\
\nonumber
&&+\gamma_{\mu} S(k-p) S(k-q)\gamma_{\nu} S(k)\Bigr),
\label{gammamunu}
\eea
and the one-loop 3-photon function $\Gamma_{\mu\nu\lambda}$ which
is given by
\bea
\Gamma_{\mu\nu\lambda}(p,-q,q-p) &=& - {e^3 \over \sqrt N} \int \!
{d^3 k\over (2\pi)^3}
{\rm tr} \Bigl(\gamma_\mu S(k-p)\gamma_\nu S(k+q-p)\gamma_\lambda S(k)
\nonumber \\
&&+\gamma_\mu S(k-p)\gamma_\lambda S(k-q)\gamma_\nu S(k)\Bigr).
\label{gammamunulambda}
\eea
Their contributions to the vacuum polarization may be written as
\be
\Delta_2\Pi_{\mu\nu} (p)=\int \! {d^3 q\over
(2\pi)^3}\Gamma_{\mu\rho (1)}(p,-q,q-p) D(q-p)G^{\rho\sigma}(q)
\Gamma_{\nu\sigma (1)}(-p,q,p-q),
\label{deltatwo}
\ee
\be
\Delta_3\Pi_{\mu\nu}(p)=\int {d^3 q\over (2\pi)^3}\Gamma_{\mu\rho\beta}
(p,-q,q-p)G^{\alpha\beta}(q-p)
G^{\rho\sigma}(q)
\Gamma_{\nu\sigma\alpha}(-p,q,p-q).
\label{deltathree}
\ee

Their null contributions to the coefficient of the CS term can be
discussed in the same spirit of the non-renormalization theorem [5]
in the weak coupling expansion. The vanishing of the corrections in
$\Delta_2\Pi_{\mu\nu}$ and $\Delta_3\Pi_{\mu\nu}$ to the coefficient
of the CS term follows from the observation that the CS term is of
order of $p$ while $\Delta_2\Pi_{\mu\nu}$ and $\Delta_3\Pi_{\mu\nu}$
are of order $p^2$ in the limit where $p\rightarrow 0$.
We can see that the gauge invariance
\be
p^\mu\Gamma_{\mu\rho (1)}(p,-q,q-p)=0,\
p^\mu\Gamma_{\mu\rho\beta}(p,-q,q-p)=0
\ee
and the analyticity of the one-loop functions yields that
\be
\Gamma_{\mu\rho (1)}(p,-q,q-p)=O(p),\
\Gamma_{\mu\rho\beta}(p,-q,q-p)=O(p)
\ee
as $p \rightarrow 0$. Similarly we also find that
\be
\Gamma_{\nu\sigma (1)}(-p,q,p-q)= O(p),\
\Gamma_{\nu\sigma\alpha}(-p,q,p-q)= O(p).
\ee
It follows from this that the integrands in Eqs.~(\ref{deltatwo}) and
(\ref{deltathree}) may be of order $p^2$. But it is yet to be
examined whether the integration over
$q$ may change the order in $p$; that is possible if the integrands
have singularities as $q\rightarrow p$.

The singular behavior of the photon propagator $G_{\alpha \beta}
(q-p)$ depends on whether $\theta$ is cancelled by $\Pi_o(0)$ which
is the one-loop corrections to $\theta$.
When $\Pi_o(0) + \theta \neq 0$, the photon propagator
$G_{\alpha\beta}(q-p)$ behaves as $q\rightarrow p$
\be
G_{\alpha\beta}(q-p) \rightarrow \left\{
\begin{array}{ll}\displaystyle
 {1\over {(e^2/4\pi)^2}} {2\over {3M}}
\ ({\rm symmetric} \ {\rm part})\\
\displaystyle
{1 \over {(e^2/4\pi)^2}} {1 \over {|q - p|}}
\ ({\rm antisymmetric}\ {\rm part}).
\end{array} \right.
\ee
When $\Pi_o(0) + \theta = 0$, the antisymmetric part of the photon
propagator vanishes and the symmetric part introduces a singularity
\be
G_{\alpha\beta}(q-p) \rightarrow {3M \over 2}{1 \over (q - p)^2}.
\ee
But the gauge invariance and the analyticity ensure that the two one-loop
photon functions, Eqs.~(\ref{gammamunu}) and (\ref{gammamunulambda}),
introduce a factor of $(q-p)^2$. Thus the
integrand in Eq.~(\ref{deltatwo}) is nonsingular where $q=p$.

On the other hand, the two one loop 2-photon 1-$\sigma$ field function
in Eq.~(\ref{deltathree}) does not introduce the factor $(q-p)^2$,
since the fermion-$\sigma$ vertex is not associated with  the gauge
invariance. Therefore if the $\sigma$ propagator $D(q-p)$ has a
singularity at $q=p$, the integration over $q$ may change
the order of $p$. Fortunately the $\sigma$ propagator $D(q-p)$ is
regular as $q\rightarrow p$
\be
D(q-p) \rightarrow {2 \pi \over M}.
\ee
The above arguments also apply to the diagrams of Fig.~6(c),
therefore they lead us to conclude that there are no corrections
to the coefficient of the CS term at $O(1/N)$.
Our explicit calculation confirms that there
are no infinite radiative corrections at $O(1/N)$.

We can extend the arguments discussed above and apply them to the
corrections from the diagrams at higher orders in $1/N$.
In general, higher order diagrams for the vacuum polarization
consist of the (fermion) one-loop functions, the internal photon
and $\sigma$-field lines. The internal lines
represent the induced propagators (See Eqs.~(\ref{auxil}),
(\ref{gmunu}).) and connect the legs
of the one-loop diagrams together all but two photon legs which are
the external ones which carry the momenta $p$ and $-p$ respectively.
Since the $1/N$ expansion respects the gauge invariance,
\be
p^{\mu_i}\Gamma_{\mu_1\dots\mu_i\dots\mu_m (n-m)}(p_1,\dots,p_i,
\dots,p_m, p_{m+1},\dots,p_n)=0
\label{pwithgamma}
\ee
as $p^{\mu_i}\rightarrow 0$ \cite{hill} where $\quad \sum^n_{j=1}
p_j=0, (i=1,\dots,m)$ and the analyticity of the one-loop functions
implies that, for $m \geq 3$,
\be
\Gamma_{\mu_1\dots\mu_i\dots\mu_m (n-m)}(p_1,\dots,p_i,\dots,p_m,
p_{m+1}, \cdots p_n) = O(p_1,\dots,p_i,\dots,p_m).
\label{magni}
\ee

The two external photon legs may be attached to the same one-loop diagram
or to two seperate one-loop diagrams. Then the
integrand for the vacuum polarization tensor contains
$\Gamma_{\mu\nu\dots (m)}(p,-p,\dots)$ in the former case and
$\Gamma_{\mu\dots (m)}(p,\dots) \Gamma_{\nu\dots (n)}(-p,\dots)$ in the
latter case. In either case, Eq.~(\ref{magni}) shows that the
one-loop functions introduce $p^2$ in the integrand.
The photon propagators may introduce singularities when their momenta
vanish. But the gauge invariance and the analyticity again assure that
the one-loop functions, corresponding to the one-loop
diagrams where the photon propagators are attached to, vanish precisely
such that the integrands are free of singularities. As we
discussed in the corrections at order $1/N$, the induced $\sigma$
propagator does not introduce any singularity in the infrared region.
Therefore we can conclude that the integrations over the internal
momenta do not change the order in $p$ and the resultant
corrections to vacuum polarization are of order $p^2$, i.e., no higher
order corrections to the CS term in the $1/N$ expansion.

However, these arguments on the vanishing radiative corrections do not
apply to the unbroken phase in which the fermions remain massless. The
absence of the analyticity in the massless case indeed results in a
correction to $\theta$ in the weak coupling expansion [7,9]. This may
happen in the unbroken phase. In this case, we can still
argue that there will be no infinite radiative correction to $\theta$
but only finite one, since the mass of the fermions would not change
the ultraviolet structure of the theory.

We find that the CS term affects the high energy behavior of the
(2+1)-dimensional Chern-Simons QED with a four-Fermi interaction. The
nature of the critical four-Fermi coupling at $O(1/N)$ is investigated
under the CS influence. The $\beta$ function around the critical
coupling is calculated at $O(1/N)$. By explicit calculation, we
find that the radiative correction to $\theta$ vanishes at $O(1/N)$ in
the dynamically broken phase of the parity symmetry. We also prove that
there is no radiative corrections to $\theta$ at higher orders
in $1/N$.

\section*{Acknowledgements}
J.~Chay is supported by Basic Science Research Institute Program,
Ministry of Education, Project No.~BSRI-92-218 and by Korea
University. D.~K.~Hong is supported in part by KOSEF and in part by
Maeji Institute at Yonsei University. T.~Lee thanks
Professor H.~S.~Song for the hospitality during his visit at the
Center for Theoretical Physics, Seoul National University. T.~Lee is
supported in part by KOSEF and in part by non-directed Korea Research
Foundation (1992).

\pagebreak

\pagebreak

\centerline{Figure Captions}

Figure 1: Feynman rules in the phase of the broken parity.

Figure 2: Forbidden diagrams.

Figure 3: Diagrams contributing to $Z$'s.

Figure 4: $N$ versus $x$.

Figure 5: Two-loop diagrams at $O(1/N)$ contributing to $Z_4$.

Figure 6: Three-loop diagrams at $O(1/N)$ contributing to $Z_4$.

\end{document}